\documentclass[rmp,aps,twocolumn]{revtex4-2}

\usepackage{graphicx}
\usepackage{bm}
\usepackage{booktabs}
\usepackage{makecell}
 
\usepackage{relsize} 
\usepackage[english]{babel}
\usepackage{hyperref}

\begin{document}
\title{\textbf{Quantum Sensors for Chemistry and Materials Science} }

\author{Piotr~Put$^{1,2}$, Arjun~Pillai$^{3}$, Xuan~Hoang~Le$^{1,3}$, Mikhail~D.~Lukin$^1$, and Hongkun~Park$^{1,3,}$}
\email{hpark@g.harvard.edu}
\affiliation{$^1$Department of Physics, Harvard University, Cambridge, Massachusetts 02138, USA.}
\affiliation{$^2$Marian Smoluchowski Institute of Physics, Jagiellonian University in Krak\'ow, 30-348 Krak\'ow, Poland}
\affiliation{$^3$Department of Chemistry and Chemical Biology, Harvard University, Cambridge, Massachusetts 02138, USA}

\begin{abstract}
The advancement of chemistry and materials science relies on transformative analytical tools which can overcome the sensitivity, spatial resolution, and throughput limitations of conventional techniques. This review explores the application of quantum sensors — specifically optically pumped magnetometers (OPMs) and nitrogen-vacancy (NV) centers in diamond — as robust platforms for molecular and materials analysis. We contrast the extreme magnetic sensitivity of macroscopic OPM ensembles with the atomic-scale resolution and multimodal capabilities of solid-state NV centers. We highlight their deployment in zero- to ultralow-field and nanoscale NMR spectroscopy, real-time reaction monitoring, and transient radical and pH detection. Furthermore, we discuss their integration into high-throughput chemical assays and non-destructive materials diagnostics, such as operando battery monitoring. With the ongoing commercialization of these technologies and advances in quantum-enhanced sensitivities, quantum sensors are poised to routinely address complex real-world analytical challenges.
\end{abstract}


\maketitle

\tableofcontents

\section{\label{sec:sec1}Introduction}

Breakthroughs in chemistry and materials science are historically catalyzed by the advent of transformative physical measurement tools. Techniques such as NMR~\cite{ernst_principles_1990}, electron and scanning probe microscopy~\cite{binnig_surface_1982}, X-ray crystallography~\cite{bragg_reflection_1913}, and advanced optical spectroscopy~\cite{demtroder_laser_2014} and microscopy~\cite{hell_breaking_1994, betzig_imaging_2006} have all fundamentally redefined our ability to characterize matter. Despite their immense utility, these conventional analytical methods are increasingly constrained by physical limits. Key challenges include bounded sensitivity and high limits of detection (LOD), a reliance on bulk-averaged information that obscures local molecular heterogeneity, insufficient spatial resolution, and difficulties in achieving high-throughput multiplexing without sacrificing precision.

To overcome these barriers, researchers are turning to the rapidly expanding domain of quantum science and technology. This field relies on the precise control of isolated quantum systems --- exploiting quantum superposition and entanglement --- to process information and interact with the physical world in ways classical systems fundamentally cannot~\cite{degen_quantum_2017}. While the best known applications involve building quantum computers and secure communication networks, these efforts face a major challenge associated with the extreme sensitivity of quantum states to environmental perturbations that destroys their delicate coherence, making it difficult to preserve quantum information. Quantum sensors present a paradigm shift that turns this fundamental vulnerability into a potential asset~\cite{degen_quantum_2017}. By deliberately exposing well-controlled quantum systems to target environments, their fragility makes them prime analytical probes~\cite{aslam_quantum_2023, doherty_nitrogen-vacancy_2013}.

Such quantum sensors provide an emerging solution to classical analytical bottlenecks, offering performance capabilities that span two physical extremes: from unmatched macroscopic \emph{sensitivity} allowing detection of minute signals~\cite{kominis_subfemtotesla_2003}, down to true atomic-scale \emph{spatial resolution}. Furthermore, they unlock entirely new metrological modalities. For instance, advanced protocols allow direct measurements of multipoint spatiotemporal correlations via covariance magnetometry~\cite{rovny_nanoscale_2022}, while the utilization of quantum entanglement and spin squeezing can bypass quantum projection noise limits entirely~\cite{sewell_magnetic_2012, rovny_multi-qubit_2025}.

This review specifically highlights the translation of this emerging class of sensors into applied tools for chemistry and materials science. Historically, deploying these sensors required deep physics expertise and complex laboratory setups. However, the field is maturing rapidly. With commercial, user-friendly versions of quantum sensors now becoming available~\cite{aslam_quantum_2023, tierney_optically_2019, bongs_taking_2019}, this review captures a crucial moment of transition, demonstrating that these robust quantum platforms are poised to be used more routinely for solving complex, real-world problems in the chemical and material sciences.

\section{\label{sec:sec2}Principles of quantum sensing}

While quantum sensing\footnote{Quantum sensors: A class of devices that exploit the coherent control of quantum systems to measure physical quantities, such as time, acceleration, or magnetic field, with high sensitivity and spatial resolution.}~\cite{degen_quantum_2017} constitutes a broad and rapidly growing subfield within the wider domain of quantum information and technology, this review focuses specifically on two of the most developed~\cite{aslam_quantum_2023} and widely utilized platforms in chemistry: optically pumped magnetometers (OPMs) and nitrogen-vacancy (NV) centers in diamond.

\begin{figure*}[btp!]
\begin{center}
\includegraphics[width = \textwidth]{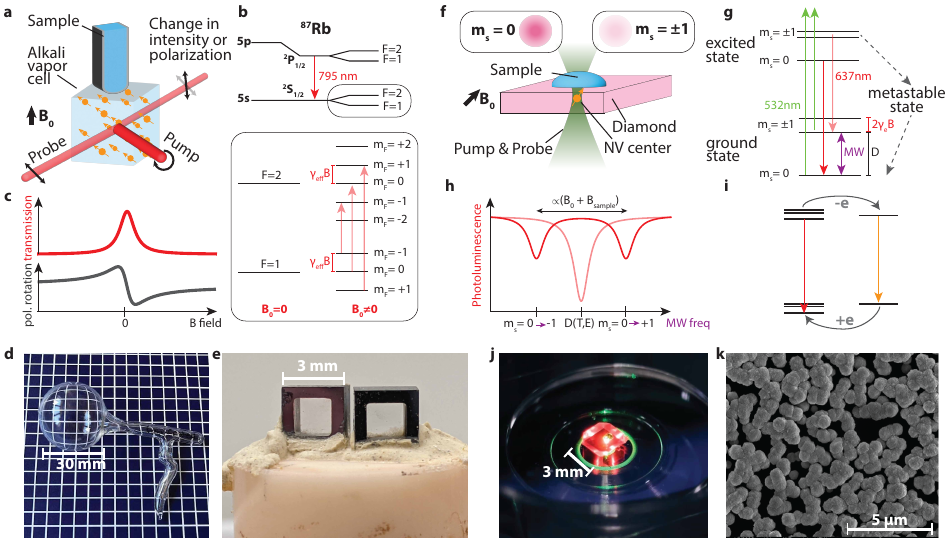}
\caption{{\bf{Quantum sensors.}}
(a) Principle of OPM-based sensing. Sample is brought close to glass cell hosting alkali-metal vapor. Atoms in the vapor are optically polarized by a pump laser beam. Sample magnetic field leads to spin evolution, changing the absorption or polarization rotation which is probed with a weak laser beam.
(b) Energy-level structure of alkali atoms (here $^{87}$Rb) used in OPMs. A laser tuned to (or near) an optical transition in the fine structure of the atom is used as a pump and probe. External magnetic field lifts the degeneracy of the energy levels within the hyperfine manifolds. The population and coherence between these Zeeman sublevels are then optically probed.
(c) Illustration of the vapor's optical properties — the transmission of on-resonance light and the polarization rotation of linearly polarized detuned light — as a function of the magnetic field. Changes are monitored to measure the magnetic field.
(d) A glass vapor cell. Large active sensing volume enables high magnetic sensitivity.
(e) Photograph of alkali-metal vapor microcells, which form the core sensing volume in many minature OPMs.
(f) Principle of NV-based sensing. NV centers are typically excited with green light, and red PL is measured. PL intensity is coupled to the electron spin state of the center ($m_s=\pm 1$ is relatively darker than $m_s=0$). 
(g) Simplified energy-level diagram of the negatively charged NV center (NV$^-$). Sensing typically relies on coherent spin dynamics within the ground state, which can be manipulated using microwaves (MWs). A non-radiative metastable state facilitates optical spin polarization and provides contrast for PL-based readout. 
(h) NV PL as a function of swept MW frequency during an ODMR measurement. Dips in the PL correspond to transitions between the ground-state spin sublevels. The position of the zero-field splitting parameter $D$ can be used to measure temperature or electric fields, while probing the Zeeman splitting allows for magnetometry.
(i) Simplified energy structure illustrating the NV$^-$ and NV$^0$ charge states. The local electrostatic environment of the NV can induce charge-state conversion, which can be optically detected due to distinct PL spectra.
(j) A bulk diamond with a high density of NV centers under green optical excitation. The characteristic red PL is visible.
(k) Scanning electron micrograph of nanodiamonds hosting NV centers. 
}
\label{fig: 1}
\end{center}
\end{figure*}

\subsection{\label{subsec:opm}Optically pumped magnetometers}

OPMs represent a class of quantum sensors that derive their performance from the collective spin dynamics of an alkali-metal vapor ensemble (\textbf{Table~\ref{tab: sensor_comparison}})~\cite{budker_optical_2007}. The core principle of operation relies on controlling and measuring these atomic spins using light (\textbf{Figure~\ref{fig: 1}a}). As illustrated by the energy-level diagram (\textbf{Figure~\ref{fig: 1}b}), a resonant pump laser optically polarizes the vapor of atoms such as $^{87}$Rb, aligning the atomic spins into a well-defined initial state. When a sample is brought into the proximity of the vapor cell (typically with a $\sim$5~mm stand-off), its magnetic signatures drive the coherent spin evolution of the sensor atoms. This spin evolution directly alters the optical properties of the atomic cloud, which is continuously read out by monitoring the resonant transmission or polarization rotation of a weak probe laser (\textbf{Figure~\ref{fig: 1}c}). Because they leverage a massive number of atoms simultaneously ($10^{11}-10^{14}$), OPMs achieve extraordinary magnetic sensitivities, reaching $0.2 - 100~\mathrm{fT}/\sqrt{\mathrm{Hz}}$ in the DC-to-kHz bandwidth at zero bias field. This extreme sensitivity makes them exceptionally well-suited for the non-invasive, high-precision characterization of bulk chemical samples at standard operating temperatures (300 -- 500 K).

This sensitivity relies on preserving the transverse spin coherence time (T$_2$) against wall collisions and inter-atomic spin-exchange. Wall-induced depolarization is mitigated by coating the cell interior with anti-relaxation hydrocarbon layers (\textbf{Figure~\ref{fig: 1}d})  or introducing inert buffer gases (e.g. N$_2$, $^4$He) to restrict alkali atoms to a slow diffusive random walk (\textbf{Figure~\ref{fig: 1}e}).

To eliminate spin-exchange decoherence, sensors often operate in the spin-exchange relaxation-free (SERF) regime~\cite{happer_spin-exchange_1973}. By combining high atomic density with a near-zero magnetic field, rapid spin-exchange collisions lock the atoms into a tightly coupled macroscopic spin state. This preserves ensemble coherence for tens to hundreds of milliseconds, ultimately enabling sub-femtotesla magnetic sensitivity~\cite{kominis_subfemtotesla_2003}.

\subsection{\label{subsec:nv}NV centers in diamond}

In contrast to the macroscopic ensemble-based nature of OPMs, NV centers in diamond offer a solid-state quantum sensing platform characterized by unprecedented spatial resolution and intimate target proximity (\textbf{Table~\ref{tab: sensor_comparison}}). An NV center is an atomic-scale point defect within the diamond lattice, consisting of a substitutional nitrogen atom and an adjacent lattice vacancy~\cite{doherty_nitrogen-vacancy_2013}. Sensing with NVs typically relies on monitoring the red fluorescence under green light illumination (\textbf{Figure~\ref{fig: 1}f}).  The energy structure of the relevant electrons constituting this defect is illustrated in \textbf{Figure~\ref{fig: 1}g}. Continuous green laser illumination (e.g., 532 nm) polarizes the defect into its $m_s=0$ ground state (\textbf{Figure~\ref{fig: 1}g}). This optical pumping occurs because the excited $m_s= \pm1$ states possess a high probability of intersystem crossing into metastable singlet states, which decay preferentially into the $m_s=0$  ground state. Furthermore, the transition of the $m_s=\pm 1$ populations via these singlet states is non-radiative. Consequently, the overall red photoluminescence (PL) is lower compared to that of the initial $m_s=0$ population, which allows for optical spin readout (\textbf{Figure~\ref{fig: 1}f}). Similar to OPMs, long coherence times are critical for high sensitivity. The NV center ground state possesses millisecond-level longitudinal relaxation times (T$_1$)\footnote{Coherence times T$_1$ and T$_2$: Time scales of quantum sensor decoherence due to spin relaxation or dephasing processes respectively, caused by environmental noise at different energy scales.}~even at room temperature. By utilizing isotopically purified diamond matrices (with $<0.01\%$ $^{13}\text{C}$), the transverse coherence time (T$_2$) can also reach millisecond levels, yielding state-of-the-art nanoscale sensitivity~\cite{rondin_magnetometry_2014, barry_sensitivity_2020}.

The core sensing mechanism with NVs relies on optically detected magnetic resonance (ODMR) (\textbf{Figure~\ref{fig: 1}h}). In this measurement, red fluorescence is monitored while the frequency of an applied microwave field is swept. When the microwave drive is resonant with the allowed transitions between the Zeeman sublevels, it drives a spin population transfer from the $m_s=0$ to the $m_s=\pm 1$ states, resulting in a detectable drop in fluorescence. In the presence of an external magnetic field, the degenerate $m_s=\pm 1$ levels experience a Zeeman shift; this splits the ODMR dips, allowing precise quantification of the local DC magnetic field (\textbf{Figure~\ref{fig: 1}h}). 
Furthermore, AC magnetic fields can be detected using dynamical decoupling\footnote{Dynamical decoupling: A periodic pulse sequence that acts as a narrow bandpass filter, decoupling the sensor from broadband environmental noise while selectively accumulating phase from a frequency-matched AC magnetic signal.}~protocols, which constructively ``lock'' the phase accumulation of the NV spin superposition state to the oscillating field. By repeating dynamical decoupling sequences at different frequencies, the spectral components of a multi-frequency signal can be reconstructed. Another sensing modality is relaxometry or decoherence spectroscopy, which probes the variance of fluctuating magnetic fields through changes in the NV T$_1$ or T$_2$ times~\cite{degen_quantum_2017}.

Beyond magnetometry, NV centers are inherently multi-modal sensors (\textbf{Table~\ref{tab: sensor_comparison}}). The zero-field splitting $D$, caused by the dipole-dipole interaction of the defect's electrons, depends on temperature, strain-induced deformation of the diamond lattice, and also local electric fields. Moreover, the charge state of the center (either $\text{NV}^-$ or $\text{NV}^0$) can be altered by nearby charge distributions (\textbf{Figure~\ref{fig: 1}i}), resulting in a shift in the NV PL spectrum that can be leveraged for all-optical sensing of local charge densities. More details about NV sensing experimental setups and protocols can be found in Refs.~\cite{schirhagl_nitrogen-vacancy_2014, bucher_quantum_2019, levine_principles_2019, scholten_widefield_2021}.

While OPMs excel in ultimate bulk magnetic sensitivity, NV centers trade this absolute sensitivity for extreme spatial localization. Their DC magnetic sensitivity typically ranges from $5~\mathrm{pT}/\sqrt{\mathrm{Hz}}$ to $1~\mathrm{\mu T}/\sqrt{\mathrm{Hz}}$~\cite{barry_sensitive_2024, sekiguchi_diamond_2024, omar_human_2026}. However, because they can be brought into nearly direct physical contact with the target system---bypassing the stand-off distances required by vapor cells---their detection volume approaches the atomic scale, ranging from atto- to nano-liters (\textbf{Table~\ref{tab: sensor_comparison}}).  Additionally, NV centers can operate robustly across a massive temperature range (0 to 600 K) and maintain their functionality even in high on-axis background magnetic fields that would otherwise saturate an OPM (\textbf{Table~\ref{tab: sensor_comparison}}). Their robust solid-state nature allows NV sensors to be deployed in highly versatile form factors tailored to specific applications. For example, they can be engineered as sensing layers in bulk diamond chips (\textbf{Figure~\ref{fig: 1}j}) or utilized as fluorescent nanodiamonds that are easily integrated into liquid samples (\textbf{Figure~\ref{fig: 1}k})~\cite{rondin_magnetometry_2014, van_der_laan_nanodiamonds_2018}.

Ultimately, the contrasting but complementary capabilities of OPMs and NV centers provide a highly versatile quantum sensing toolkit. While OPMs push the absolute limits of macroscopic magnetic sensitivity for bulk sample analysis, NV centers offer unprecedented spatial resolution, environmental robustness, and multi-modal sensing capabilities at the nanoscale. In the following sections, we examine how these distinct physical advantages are being actively translated into transformative analytical techniques across the chemical and material sciences, enabling everything from nanoscale NMR and non-destructive battery diagnostics to the real-time tracking of transient radical species.

\begin{table*}[t]
\small
\setlength{\tabcolsep}{4pt}
\caption{\textbf{Comparison of OPM and NV center key properties and figures of merit.}}
\label{tab: sensor_comparison}
\begin{tabular}{lll}
\toprule
\toprule

\textbf{Property/ FOM} &
\textbf{OPMs} &
\textbf{NV centers} \\
\midrule

\textbf{Sensor type} &
spin ensemble &
single spin or ensemble \\

\textbf{Measured quantity} &
magnetic field/gradient &
\begin{tabular}[t]{@{}l@{}}
magnetic field/field gradient, temperature,\\
electric field/charge, strain/pressure
\end{tabular} \\

\textbf{Physical size} & mm to cm (vapor cell + optics) & atomic scale (defect); mm (bulk chip) \\

\textbf{Spatial resolution} & $\sim$1 mm to 1 cm & $\sim$10 nm (single NV) to 1 \textmu m (ensemble) \\

\textbf{Detection volume} & \textmu L to mL & aL to nL \\

\textbf{Target proximity} & macroscopic ($\sim$mm standoff) & atomic (can be in direct contact) \\

\textbf{Operating temperature} & 300 - 500 K & 0 - 600 K \\

\begin{tabular}[t]{@{}l@{}}
\textbf{Background field range}
\end{tabular} &
mode dependent (100 nT to $\sim$100 \textmu T) &
\begin{tabular}[t]{@{}l@{}}
on-axis: high (unaffected by Tesla-scale fields)\\
off-axis: $< 10$ mT
\end{tabular} \\


\textbf{Magnetic bandwidth} & DC - kHz & DC - GHz \\

\textbf{DC magnetic sensitivity\footnote{Magnetic field sensitivity: The smallest magnetic field detectable by coherent $T_2$ sensing within 1 s of averaging, in units of $\mathrm{T}/\sqrt{{\mathrm{Hz}}}$; or, the smallest mean square magnetic noise detectable by $T_1$ relaxometry in a 1 Hz band within 1 s of averaging, in units of $\mathrm{T}/\sqrt[4]{{\mathrm{Hz}}}$.}} & \thead[l]{0.16 - 100 fT/$\sqrt{\mathrm{Hz}}$ \footnotemark} & \thead[l]{5  pT/$\sqrt{\mathrm{Hz}}$ - 1 \textmu T/$\sqrt{\mathrm{Hz}}$  \footnotemark}\\

\textbf{AC magnetic sensitivity} & 
\begin{tabular}[t]{@{}l@{}}
kHz frequencies: 0.2 - 100 fT/$\sqrt{\mathrm{Hz}}$ \footnotemark
\end{tabular} &
\begin{tabular}[t]{@{}l@{}}
kHz-MHz frequencies:
200 fT/$\sqrt{\mathrm{Hz}}$ - 1 \textmu T/$\sqrt{\mathrm{Hz}}$ \footnotemark\\
few-GHz noise:
10 pT$_{rms}$/$\sqrt[4]{\mathrm{Hz}}$ - 100 nT$_{rms}$/$\sqrt[4]{\mathrm{Hz}}$  \footnotemark
\end{tabular}\\


\bottomrule
\bottomrule

\end{tabular}

\footnotetext[2]{\cite{kominis_subfemtotesla_2003, dang_ultrahigh_2010, sheng_subfemtotesla_2013}}~
\footnotetext[3]{\cite{barry_sensitive_2024, sekiguchi_diamond_2024, omar_human_2026}}~
\footnotetext[4]{\cite{lee_subfemtotesla_2006}}~
\footnotetext[5]{\cite{barry_sensitive_2024, gao_high_2023}}~
\footnotetext[6]{\cite{andersen_electron-phonon_2019, rovny_nanoscale_2024}}


\end{table*}

\section{\label{sec:sec3}Chemical analysis with quantum sensors}

The outstanding \emph{sensitivity} and \emph{resolution} of quantum sensors enable new modalities of chemical analysis in regimes inaccessible to conventional techniques. This section surveys several of these new modalities, including NMR at zero- to ultralow-field or of picoliter volumes, direct monitoring of chemical reactions within bulk volumes or at interfaces, and probing radical formation and pH with nanoscale resolution.

\subsection{\label{subsec:chemicalIdentification}Chemical identification via high-resolution NMR}

\begin{figure*}[btp!]
\begin{center}
\includegraphics[width=\textwidth]{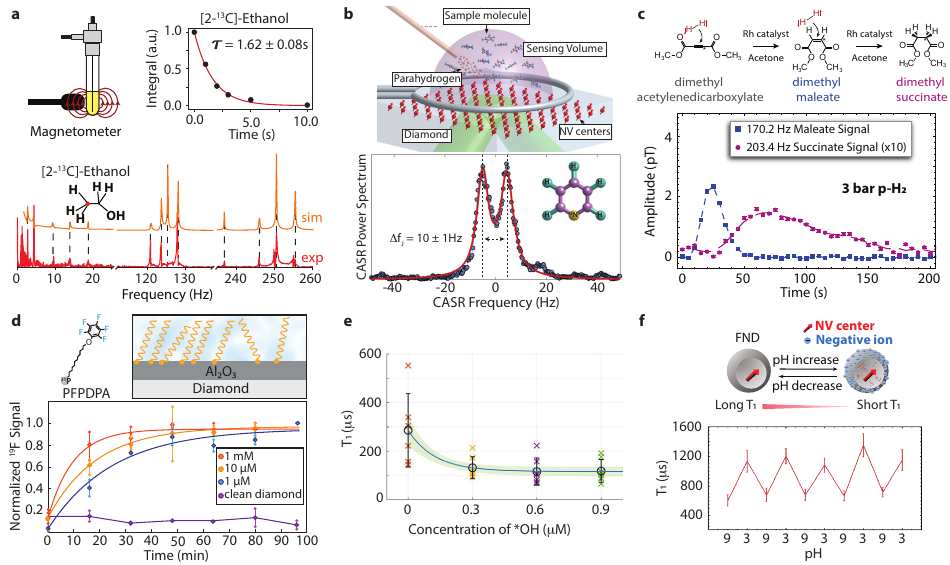}
\caption{{\bf{Chemical identification using quantum sensors.}}
(a) NMR chemical fingerprinting using OPM detection. A compact OPM (top left) records the NMR signal from a liquid sample hyperpolarized with parahydrogen under zero magnetic field, yielding a chemically specific zero-field spectrum (bottom) and enabling relaxometry measurements (top right). The spectrum reflects J-couplings between nuclear spins in the investigated molecule (2-$^{13}$C ethanol). Adapted from Ref.~\cite{van_dyke_relayed_2022}.
(b) Chemically specific proton NMR with NVs. High spectral resolution and micron-scale spatial resolution is achieved via hyperpolarization of $^{15}$N-labeled pyridine. $\sim$10~Hz J-coupling between $^{15}$N and $^1$H spins is easily resolvable. Adapted from Ref.~\cite{arunkumar_micron-scale_2021}.
(c) Real-time chemical monitoring by OPM-ZULF NMR. Formation of reaction products from partial hydrogenation (maleate) and full hydrogenation (succinate) is monitored in real time as the reaction with parahydrogen proceeds. Adapted from Ref.~\cite{burueva_chemical_2020}.
(d) Detection of self-assembled monolayer formation using surface-sensitive NMR with $\sim$5-nm deep NVs. $^{19}$F detection allows for real-time monitoring of PFPDPA binding to Al$_2$O$_3$ on the diamond surface. Adapted from Ref.~\cite{liu_surface_2022}.
(e) Radical detection with NVs in nanodiamonds. Higher *OH concentration exponentially reduces NV T$_1$. Adapted from Ref.~\cite{perona_martinez_nanodiamond_2020}.
(f) Reversible pH sensing with NVs with carboxyl-functionalized nanodiamonds. At higher pH, the carboxyl groups are converted to negatively charged carboxylate ions, impacting NV T$_1$. Adapted from Ref.~\cite{fujisaku_ph_2019}.
}

\label{fig: 2}
\end{center}
\end{figure*}

Nuclear magnetic resonance (NMR) is a powerful tool for studying molecular structure and dynamics. Quantum sensors offer a low-footprint alternative to high-field magnets, with recent efforts focused on achieving the sensitivity and spectral resolution required for chemically specific NMR.

While pushing to higher magnetic fields improves the spectral resolution and chemical specificity of NMR, particularly for large molecules, the opposite regime — zero- to ultralow-field (ZULF) NMR\footnote{Zero- to ultralow-field (ZULF) NMR: A regime of nuclear magnetic resonance performed without strong magnetic background fields, where spectra are dominated by spin-spin couplings rather than chemical shifts.}~— has emerged in recent years as a promising alternative, allowing for high-resolution spectroscopy without bulky superconducting magnets~\cite{barskiy_zero-_2025}. In this regime, NMR spectra are no longer governed by chemical shifts, but instead dominated by scalar spin-spin ($J$) couplings, providing a unique spectral fingerprint of molecular structures. OPMs are the main enabler of this development, as they offer state-of-the-art magnetic sensitivity in a compact, low-power, and cost-effective integrated sensing package. Early foundational work~\cite{ledbetter_zero-field_2008, ledbetter_near-zero-field_2011} showed that OPMs possess the requisite sensitivity to detect nuclear spin precession at near zero fields, soon followed by a demonstration of high-resolution zero-field $J$-spectroscopy of complex aromatic compounds~\cite{blanchard_high-resolution_2013}. Because ZULF NMR does not suffer from the magnetic susceptibility broadening that plagues high-field measurements in heterogeneous media, it is well-suited for studying complex samples. Building on this, recent advancements have extended OPM-ZULF NMR techniques to the spectroscopy of small biomolecules~\cite{put_zero-_2021} and the relaxometry of various chemical and biological fluids~\cite{alcicek_zero-_2023}. Furthermore, these techniques have enabled the construction of a natural-abundance $^{13}$C ZULF NMR library~\cite{andrews_dft-assisted_2026}, demonstrating the growing analytical utility of OPM-NMR in chemical and biomedical research.

Despite these capabilities, a major limitation of ZULF NMR has been the vanishingly small sample thermal spin polarization at zero field. To circumvent this, OPM detection is frequently synergized with hyperpolarization\footnote{Hyperpolarization: A class of techniques used to enhance nuclear spin polarization by orders of magnitude compared to thermal equilibrium.}~methods. The pioneering application of parahydrogen-induced polarization (PHIP) to zero-field NMR~\cite{theis_parahydrogen-enhanced_2011}, alongside the subsequent development of signal amplification by reversible exchange (SABRE) at zero field~\cite{theis_zero-field_2012}, yielded orders-of-magnitude enhancement in signal-to-noise ratio (SNR). The analytical scope of hyperpolarized OPM-ZULF NMR continues to rapidly expand through novel polarization transfer and generation strategies~\cite{put_detection_2023}. The recent implementation of relayed hyperpolarization techniques allows for the sensitization of a much broader array of molecular targets (\textbf{Figure~\ref{fig: 2}a})~\cite{van_dyke_relayed_2022}. Furthermore, the integration of OPM-NMR with photochemically induced dynamic nuclear polarization (Photo-CIDNP) offers an effective, light-driven pathway to overcome thermal polarization limits~\cite{chuchkova_magnetometer-detected_2023}. Together, these advanced hyperpolarization strategies cement OPM-NMR as a versatile tool for chemical analysis.

While OPMs excel at probing bulk samples, NV centers provide the spatial resolution required to extend magnetic resonance to picoliter volumes and complex interfaces~\cite{allert_advances_2022, mzyk_diamond_2022, budakian_roadmap_2024}. One limitation, however, is that the spectral resolution of AC magnetometry with NV centers is typically limited by the coherence time of the NV electronic spins, precluding the detection of ppm-level chemical shifts. As a solution to this, Refs.~\cite{glenn_high-resolution_2018, schmitt_submillihertz_2017, boss_quantum_2017} introduced the coherently averaged synchronized readout (CASR) protocol, where the sensor coherence-limited pulse sequences are synchronized with a phase coherent target signal to achieve spectral resolutions limited only by the stability of an external clock. This allowed chemical-shift and J-coupling resolved NMR with NV centers, with the limitation that the coherent averaging measured only the thermally polarized subset of nuclear spins in the sample~\cite{glenn_high-resolution_2018}.

Similar to OPM-NMR, to improve NV-NMR sensitivity, CASR can be integrated with hyperpolarization techniques to enhance thermal polarization by orders of magnitude. Ref.~\cite{bucher_hyperpolarization-enhanced_2020} utilized Overhauser dynamic nuclear polarization (DNP) to measure NMR of $\sim$50 femtomoles of $t$-BuOD in picoliter volumes of D$_{2}$O, and Ref.~\cite{arunkumar_micron-scale_2021} used SABRE for micron-scale NMR of millimolar concentration of pyridine in methanol (\textbf{Figure~\ref{fig: 2}b}). Other demonstrations of chemical-shift resolved NV-NMR involve operation at high magnetic field (3T)~\cite{aslam_nanoscale_2017} or at ZULF conditions, enhanced by SABRE~\cite{omar_zero-_2026}.

Beyond the detection of spin-1/2 nuclei, quantum sensors overcome the severe sensitivity bottlenecks inherent to low-frequency nuclear quadrupole resonance (NQR) without requiring external magnetic fields. For macroscopic chemical samples, radio-frequency OPMs provide exceptional sub-femtotesla sensitivity to directly detect weak NQR signatures~\cite{lee_subfemtotesla_2006}. In the complementary microscopic regime, NV centers offer the unprecedented spatial resolution required to probe quadrupolar interactions at the nanoscale. For instance, single NVs have successfully detected the quadrupolar signatures of $\sim$30 nuclei within an atomically thin lattice~\cite{lovchinsky_magnetic_2017}. Together, these modalities enable structural characterization of quadrupolar compounds across various length scales, from bulk powders to the two-dimensional limit.

\subsection{\label{subsec:reactionMonitoring}Chemical reaction monitoring}

Monitoring reaction progress involves tracking concentrations or some other physical parameters throughout the process. The aforementioned high-resolution NMR techniques, as well as T$_1$-relaxometry and multi-parameter sensing with NVs, enable monitoring chemical reactions in real-world conditions. 

The unique operating regime of ZULF NMR, facilitated by the high sensitivity of OPMs, has opened entirely new avenues for real-time chemical reaction monitoring. Because measurements are performed without a strong magnetic background, ZULF NMR is inherently well-suited for tracking reactions in environments that are typically inaccessible or challenging for conventional high-field NMR. For example, OPM-ZULF NMR has been employed to observe complex enzymatic reactions in situ, providing a non-destructive analytical window into biocatalytic processes under natural, field-free conditions~\cite{eills_enzymatic_2023}. Furthermore, the low-frequency nature of ZULF NMR signals allows them to penetrate conductive materials, which is impossible in high-field NMR due to the radio-frequency skin effect. This enables direct monitoring of chemical reactions and heterogeneous catalysis (\textbf{Figure~\ref{fig: 2}c}) occurring inside sealed metal containers and realistic industrial-grade reactors~\cite{burueva_chemical_2020}.

While OPMs excel at probing bulk reaction dynamics in such challenging macroscopic environments, NVs provide a complementary approach for monitoring reaction kinetics at the nanoscale and directly at solid-liquid interfaces. Utilizing NV-based surface NMR, researchers have successfully tracked self-assembled monolayer formation in real time (\textbf{Figure~\ref{fig: 2}d}), enabling the direct study of chemical reaction kinetics occurring precisely at the diamond surface~\cite{liu_surface_2022}. Another powerful application of NV centers in reaction monitoring leverages their sensitivity to fluctuating local magnetic fields caused by paramagnetic species. This sensitivity makes them well-suited for tracking dynamics of electron-transfer and redox reactions. For example, dynamic quantum sensing has been utilized to monitor the conversion of low-spin to high-spin Fe$^{3+}$ by measuring the resulting changes in NV ODMR contrast. This approach has achieved LOD of just 10 attomoles when imaging paramagnetic gadobutrol contrast agents~\cite{radu_dynamic_2020}. Similarly, nanodiamond-based T$_1$ relaxometry provides a robust probe for tracking the interconversion of diamagnetic copper and paramagnetic Cu$^{2+}$ species during Fenton-like reactions in the presence of H$_2$O$_2$~\cite{padamati_insight_2022}. Together, these nanoscale measurements highlight the versatility of NV centers for probing complex chemical processes in highly localized environments.

Beyond tracking the conversion of reactants to products, quantum sensors can serve as a powerful tool for probing molecular dynamics. The shape of a ZULF NMR spectrum is sensitive to the rates of chemical exchange, which can dynamically alter the effective J-coupling network of a molecule. This sensitivity was demonstrated in studies of aqueous urea, where varying the proton exchange rate allowed researchers to effectively engineer the observed spin topology and extract kinetic parameters based on OPM detection~\cite{alcicek_zero-field-urea_2021}. More recently, these capabilities have been expanded to encompass the zero-field $J$-spectroscopy of quadrupolar nuclei. Chemical exchange modulates the complex spin-spin interactions between spin-1/2 and quadrupolar isotopes. By analyzing these modulations, ZULF NMR provides distinct insights into molecular structure and exchange phenomena, which are often obscured by rapid quadrupolar relaxation at high magnetic fields~\cite{picazo-frutos_zero-field_2024}.

Complementarily, the nanoscale resolution accessible by NV centers allows for pushing molecular dynamics spectroscopy to the interfacial level. For example, Ref.~\cite{zheng_probing_2025} utilized sensing with single NV centers and local electron injection with scanning probe microscopy to study the dissociation of interfacial water. NV sensing here allowed for detection of the highly reactive intermediates by double electron-electron resonance. Additionally, nanoscale $^1$H-NMR signal of the reaction species allowed direct measurements of the diffusion rates of both the intact water molecules and the hydroxide reaction products, revealing that interfacial hydroxide diffuses significantly faster than the surrounding water.

\subsection{\label{subsec:radicalFormation}Detecting radical formation and pH changes}

Free radicals are reaction intermediates which drive chemical and physiological processes such as bacterial responses to antibiotics, viral immune responses, and cellular aging. However, their extreme reactivity and fleeting lifespans make direct nanoscale observation difficult using conventional techniques. NVs overcome this sensing bottleneck by detecting transient radicals through two distinct physical mechanisms. T$_1$ relaxometry exploits the acceleration of the NV spin relaxation rate due to magnetic noise from target unpaired electrons~\cite{mzyk_relaxometry_2022}. Alternatively, radicals can be optically detected via radical-induced changes to the diamond surface chemistry, which alter the NV charge state and thus PL spectrum~\cite{ninio_high-sensitivity_2021}.

Early implementations validated these NV sensing mechanisms by tracking fundamental, well-characterized chemical processes. For instance, Ref.~\cite{barton_nanoscale_2020} covalently coupled nitroxide radicals to polymer-coated nanodiamonds, utilizing T$_1$ relaxometry to monitor their reduction to diamagnetic hydroxylamine during ascorbic acid oxidation with a resolution of $\sim$10 spins. Similarly, T$_1$ relaxometry was deployed to quantify hydroxyl radicals generated via hydrogen peroxide photolysis and Fenton/Haber-Weiss reactions in biologically relevant media containing salts and proteins (\textbf{Figure~\ref{fig: 2}e})~\cite{perona_martinez_nanodiamond_2020}. Complementary to relaxometry, Ref.~\cite{ninio_high-sensitivity_2021} achieved high-sensitivity ($11~\text{nM}/\sqrt{\text{Hz}}$) hydroxyl radical detection by leveraging the fluorescence contrast between different NV charge states (NV$^-$/NV$^0$) in bulk diamond. 
Building upon these demonstrations, researchers have adopted NV sensing to track radical dynamics in live biological microenvironments. For example, Ref.~\cite{norouzi_relaxometry_2022} utilized nanodiamond T$_1$ imaging to map localized free radical bursts from single bacteria treated with antibiotics and UV irradiation.

Similar to sensing free radicals, NV centers can detect changes in pH or redox potential via T$_1$ relaxometry or charge-state switching. For relaxometry, early protocols utilized spin-labeling to induce stochastically fluctuating magnetic fields~\cite{ermakova_detection_2013, steinert_magnetic_2013, sushkov_all-optical_2014}. Building on this, Ref.~\cite{rendler_optical_2017} utilized nanodiamonds coated with NV-polymer-Gd$^{3+}$ complexes, where lowering the pH or raising the redox potential cleaved the polymer linkers and released the paramagnetic Gd$^{3+}$ ions away from the diamond surface, subsequently increasing the NV T$_1$. While sensitive over a broad pH range, this cleavage is irreversible, rendering the sensors single-use. To achieve reusable, continuous monitoring, Refs.~\cite{fujisaku_ph_2019, cheng_all-fiber_2024} relied on intrinsic environmental ions rather than sacrificial spin labels. By coating nanodiamonds with a pH-sensitive ionic layer, they observed that localized charge noise modulated the NV T$_1$. For example, at high pH, surface carboxyl groups were converted to negatively charged carboxylate ions, significantly reducing the NV T$_1$ (\textbf{Figure~\ref{fig: 2}f}, top panel). This mechanism is fully reversible (\textbf{Figure~\ref{fig: 2}f}, bottom panel) and can be tuned to different functional pH ranges by altering the chemical coating (e.g., using polycysteine)~\cite{fujisaku_ph_2019}.

Beyond relaxometry, these pH-sensitive polymer coatings can also directly modulate the NV charge state. At higher pH, negatively ionized surface groups preferentially stabilize the NV$^-$ state, which is read out all-optically as a redshift in the bulk fluorescence spectrum~\cite{petrakova_charge-sensitive_2015, raabova_diamond_2019, sow_high-throughput_2020}. Applying this same principle, researchers have demonstrated that charge-state–dependent fluorescence can reliably detect dynamic changes in local electrochemical potential, establishing NV centers as robust nanoscale sensors compatible with complex electrolytic systems~\cite{karaveli_modulation_2016, fulton_probing_2024}.

\section{\label{sec:sec4}Chemical assays with quantum sensors}

Chemical assays are essential for diagnostics and process monitoring and require quantitative high-specificity and high-sensitivity methods. Quantum sensors leverage the sensitive detection techniques described in Section~3 to address these needs, achieving chemical specificity through high spectral resolution or via selective binding of sensors to target molecules. This section surveys quantum-sensor-based approaches for analyte detection and mass-limited, high-throughput analysis.

\subsection{\label{subsec:analyteDetection}Analyte detection}

\begin{figure*}[btp!]
\begin{center}
\includegraphics[width=\textwidth]{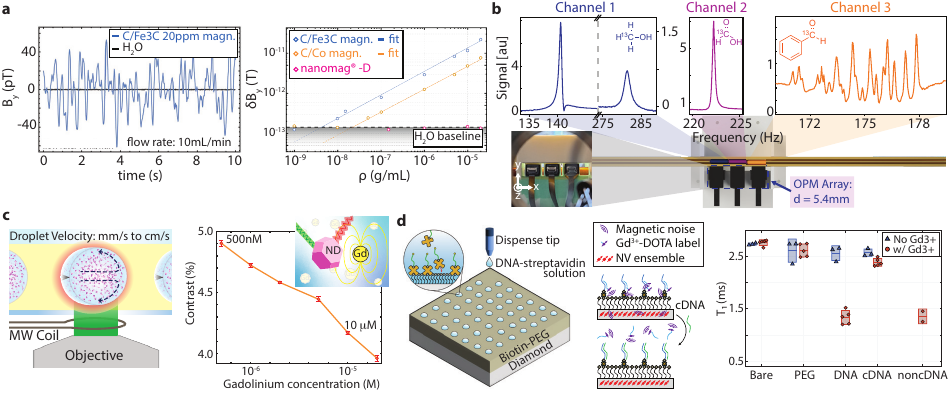}
\caption{{\bf{Chemical assays and high-throughput analysis}}
(a) In-line detection of MNPs in flowing liquids. The magnetic field generated by passing MNPs is detected by an OPM (left). The corresponding magnetic-field variance plotted as a function of particle concentration (right) exhibits linear scaling over several orders of magnitude, with a LOD of $\sim$10 ppb. Adapted from Ref.~\cite{bougas_nondestructive_2018}. 
(b) Multiplexed chemical detection using an OPM array. ZF NMR signals from different molecular species are recorded at distinct spatial locations, enabling simultaneous measurement of multiple samples. Adapted from Ref.~\cite{andrews_sensitive_2025}.
(c) High-throughput paramagnetic ion sensing. Microfluidics are integrated with confocal microscope to image nanodiamonds in flowing droplets. Lock-in detection on the droplets enables measuring ODMR contrast as a function of Gd$^{3+}$ concentration for quantitative characterization. Adapted from Ref.~\cite{sarkar_high-precision_2024}.
(d) DNA microarray on diamond. Diamond surface functionalization with biotin-PEG-silane enables binding Gd$^{3+}$-labeled single-stranded DNA targets proximal to NV layer, resulting in lower NV T$_1$. The spin label can be displaced by a controlled complementary DNA (cDNA) strand, thus restoring NV T$_1$ and enabling detection of the target. Adapted from Ref.~\cite{chi-duran_quantum_2025}. 
}
\label{fig: 3}
\end{center}
\end{figure*}

NV-based assays require robust diamond surface functionalization (e.g., Al$_2$O$_3$ deposition and PEG/PEI passivation) to enable selective target binding while preserving near-surface NV coherence~\cite{liu_surface_2022, xie_biocompatible_2022, rodgers_diamond_2024}. Many experiments also spin-label the targets to produce strong magnetic signals detectable by the NVs via ODMR-imaging or T$_1$ relaxometry~\cite{lu_magnetically_2023, zhang_patterning_2025, zalieckas_quantum_2024}, enabling detection of trace biomolecules. For example, Ref.~\cite{zalieckas_quantum_2024} demonstrated the detection of $\sim$23 microRNA molecules complexed with paramagnetic Mn$^{2+}$ labels. Furthermore, by combining these versatile polymer functionalization strategies with single shallow NVs in engineered diamond nanopillars, T$_1$ relaxometry of biomolecular interactions can be pushed to nearly the single-molecule level~\cite{li_quantum_2025}.

Alternatively, some NV spin-based protocols utilize magnetic nanoparticles (MNPs) instead of ions as labels~\cite{chen_digital_2023, kayci_multiplexed_2021, atallah_rapid_2022}. For instance, Ref.~\cite{kayci_multiplexed_2021} performed a DNA assay using NV-ODMR imaging of target molecules labeled with magnetic nanobeads. DNA-diamond interfacing was achieved by pressing a diamond chip onto a hydrogel microstructure array incorporated with the target DNA already bound to the capture DNA. This technique is sensitive to single-base mismatches between the capture and target DNA, potentially reaching $\sim$100 attomolar LOD. Ref.~\cite{atallah_rapid_2022} detected the cytokine Interleukin-6 in real COVID patient plasma samples by labeling a target protein with two nanobeads with distinct magnetic signatures. Only co-detection of these two beads indicate detection of the target, eliminating the need to distinguish bound/unbound beads. This method achieved improved speed and comparable accuracy compared to commercial Luminex assays.

Additionally, the NV charge state has been utilized for biomolecule detection. For example, negatively charged DNA molecules immobilized on PEI-functionalized diamond surfaces prepare an NV$^-$ population proportionally to the DNA concentration, quantifiable by fitting the redshifted PL spectrum, with estimated sensitivity down to $\sim$100 picomole~\cite{krecmarova_label-free_2021}.

\subsection{\label{subsec:massLimited}Mass-limited / high-throughput sample analysis}

Beyond molecular selectivity, achieving low LOD and high throughput is critical. This has driven the integration of quantum sensors with parallelized microfluidic architectures and the push towards label-free NV-NMR.

The unparalleled spatial resolution of NV centers has fundamentally redefined the physical limits of magnetic resonance. Bypassing the macroscopic volume requirements of conventional techniques, NV centers have successfully scaled NMR spectroscopy down from picoliter microfluidic volumes and 2D correlation spectroscopy~\cite{kehayias_solution_2017, smits_two-dimensional_2019} to single-protein analysis~\cite{staudacher_nuclear_2013, lovchinsky_nuclear_2016}. To overcome the sensitivity barriers at these small volumes, NV sensing has been combined with hyperpolarization techniques such as DNP and SABRE, which has yielded LOD on the femtomole scale for trace chemical analysis~\cite{arunkumar_micron-scale_2021,bucher_hyperpolarization-enhanced_2020}.

To apply these nanoscale noise spectroscopy techniques to dynamic biological fluids, researchers must overcome rapid molecular diffusion, which restricts detection time and broadens spectral lines. This barrier has been successfully addressed through spatial confinement strategies, such as trapping liquid analytes within tunable metal-organic frameworks grown directly on the diamond surface~\cite{liu_using_2022}. Beyond NMR, the extreme localized sensitivity of NV centers enables ambient electron spin resonance spectroscopy of individual paramagnetically labeled proteins, opening a window into molecular motion and conformational dynamics which are averaged out in bulk measurements~\cite{shi_single-protein_2015}.

OPMs provide complementary capabilities for mass-limited and high-throughput bulk analysis. In nanoparticle sensing, OPMs enable the nondestructive, in-line detection of MNPs at sub-picomolar concentrations within complex flowing fluids (\textbf{Figure~\ref{fig: 3}a})~\cite{bougas_nondestructive_2018}, as well as the dynamic monitoring of their clustering~\cite{everaert_monitoring_2023}. To probe even smaller volumes, OPMs coupled with high-resolution magnetic flux guides can reliably detect microscopic droplets directly within microfluidic channels~\cite{jofre_optically_2023}. Furthermore, the intrinsic scalability of OPM technology has enabled the deployment of atomic magnetometer arrays for multichannel ZULF NMR, allowing simultaneous screening of multiple samples (\textbf{Figure~\ref{fig: 3}b})~\cite{andrews_sensitive_2025}.

Diamond solid-state sensors have similarly been integrated into versatile, high-throughput microfluidic platforms. Experimental architectures have been developed for real-time widefield NMR and relaxometry microscopy~\cite{fujiwara_diamond_2023, allert_microfluidic_2022, briegel_optical_2025}, and the rapid encapsulation and detection of paramagnetic species in flowing microdroplets (\textbf{Figure~\ref{fig: 3}c})~\cite{sarkar_high-precision_2024}. Beyond flow channels, planar diamond surfaces engineered into multiplexed microarrays provide a highly parallelized framework for quantum biosensing, utilizing the strategies described in Section~4.1 (\textbf{Figure~\ref{fig: 3}d})~\cite{chi-duran_quantum_2025}. Perhaps the most impactful clinical translation of this high-throughput capability is the integration of fluorescent nanodiamonds into point-of-care lateral flow assays. Spin-dependent NV fluorescence effectively filters out the strong biological autofluorescence that limits conventional assays~\cite{miller_spin-enhanced_2020, hui_magnetically_2021}, allowing this platform to achieve extraordinary detection limits. Consequently, it has been successfully deployed to detect various viral and bacterial pathogens, including Dengue, tuberculosis, and SARS-CoV-2~\cite{le_spin-enhanced_2022, le_fluorescent_2024, wei-wen_hsiao_fluorescent_2022, thomas_decruz_quantum-enhanced_2025}.

\section{\label{sec:sec5}Materials analysis with quantum sensors}

Quantum sensors can probe a diverse array of materials, from solid-state samples to biological nanoparticles. Unlike techniques which provide only relative contrast, such as magnetic force microscopy or optical/X-ray circular dichroism, quantum sensors enable quantitative magnetic field measurements. Additionally, their non-invasive nature is well-suited to studying assembled devices under working conditions. This section highlights applications in probing magnetic order, phase transitions, and battery electrochemical environments.

\subsection{\label{subsec:magneticMaterials}Magnetic materials}

\begin{figure*}[btp!]
\begin{center}
\includegraphics[width=\textwidth]{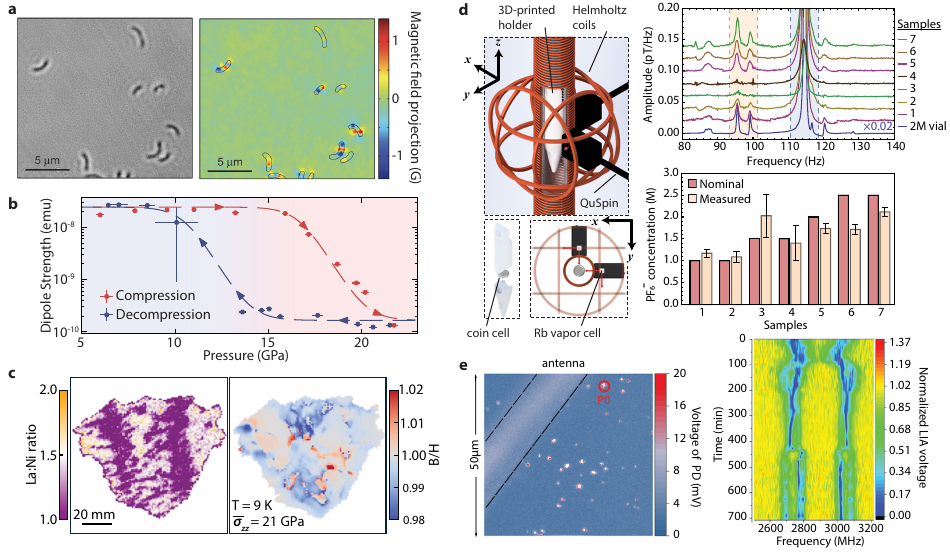}
\caption{{\bf{Quantum sensors for studying materials}}
(a) Widefield optical (left) and magnetic (right) imaging of MTB. MNPs inside magnetosomes of MTB produce stray fields detectable by NVs in the diamond chip on which the bacteria are placed. Adapted from Ref.~\cite{le_sage_optical_2013}.
(b) Probing structural phase transitions in iron. Under high pressure, iron undergoes $\alpha$ → $\epsilon$ phase transition from bcc to hcp crystal structure, leading to a change of effective dipole field the sample produces. Using NV centers in a DAC, this dipole field can be extracted by measuring the full vector magnetic field profile. Adapted from Ref.~\cite{hsieh_imaging_2019}.  
(c) Correlating local stoichiometric composition with superconductivity in high-$T_C$ nickelate superconductor La$_3$Ni$_2$O$_7$. The regions with higher Ni concentration, as determined by energy-dispersive X-ray spectroscopy (left), correspond to the regions having no diamagnetic response, and thus no Meissner effect, as measured by NV centers in a DAC (right). Adapted from Ref.~\cite{mandyam_uncovering_2026}.
(d) Measurement of electrolyte composition in batteries. The NMR signal from a coin-cell battery (left) is detected using compact OPMs yielding chemically specific zero-field spectra for various samples (top right). The amplitude of the NMR signal characteristic of PF$_6^{-}$ is used to quantify the electrolyte concentration in a non-destructive manner (bottom right). Adapted from Ref.~\cite{fabricant_enabling_2025}. 
(e) In situ measurement of electrodes in a battery. Nanodiamonds are embedded in an Fe$_3$O$_4$ electrode (left). ODMR of an embedded nanodiamond shows changes in local magnetic fields which can be attributed to the creation or depletion of paramagnetic and ferromagnetic species throughout the electrochemical evolution of the battery (right). Adapted from Ref.~\cite{liu_operando_2024}. 
}
\label{fig: 4}
\end{center}
\end{figure*}

Quantum sensors uniquely probe complex magnetic nanomaterials with high spatio-temporal resolution. Building upon early NV magnetometry~\cite{maze_nanoscale_2008, balasubramanian_nanoscale_2008} and nanoscale ferromagnetic domain imaging~\cite{rondin_nanoscale_2012, maletinsky_robust_2012}, DC magnetic imaging with NV ensembles directly maps the spatial distribution of magnetic moments. For example, this approach was used to resolve internal chains of magnetosomes in live magnetotactic bacteria (MTB)~\cite{le_sage_optical_2013, feng_-situ_2025} (\textbf{Figure~\ref{fig: 4}a}). Furthermore, by measuring fields along different NV crystallographic axes, one can obtain complete vector profiles. In the biominerals of chiton teeth, this vector profiling revealed unexpected long-range magnetic order arising from either direct dipolar coupling or spin alignment by the underlying organic matrix~\cite{mccoey_quantum_2020}.

Beyond static imaging, quantum sensors can non-invasively capture dynamic magnetic behaviors and collective population changes across multiple timescales. OPMs can detect deviations from exponential relaxation dynamics (on the order of seconds) to quantify the dispersion and rotational damping of MTB magnetic moments~\cite{ruiz_magnetotactic_2025}. OPMs also enable real-time monitoring of biophysical changes, such as magnetosome increases during incubation~\cite{feng_-situ_2025} or population decreases during solution concentration via settling and evaporation~\cite{ruiz_magnetotactic_2025}. Complementarily, NV relaxometry and noise spectroscopy can probe magnetic fluctuations at the nanoscale~\cite{rovny_nanoscale_2024, xue_magnon_2026}, allowing the identification of specific paramagnetic species and their distributions during MNP biomineralization~\cite{mccoey_quantum_2020}.

These capabilities extend seamlessly to solid-state and geological materials for precise thermomagnetic characterization. Ultrahigh-sensitivity OPMs can continuously monitor the remnant magnetization of weakly magnetic minerals as a direct function of temperature~\cite{dang_ultrahigh_2010}. Additionally, widefield quantum diamond microscopy (QDM) can provide high-throughput imaging over relatively large fields of view ($\sim$1~mm$^2$) with an area-normalized sensitivity of $\sim$20~\textmu T$\cdot$\textmu m$/\sqrt{\mathrm{Hz}}$. Capable of detecting highly localized weak magnetic moments down to $\sim$10$^{-17}$~A$\cdot$m$^2$~\cite{glenn_micrometerscale_2017, fu_highsensitivity_2020}, QDM has been used to image ferromagnetic grains in meteorites and speleothems~\cite{glenn_micrometerscale_2017, fu_high-resolution_2021, fu_pinpointing_2023, hess_investigating_2024}. By analyzing the distributions of these particles, researchers can extract sub-annual historical records of environmental conditions, such as precipitation and floods.

\subsection{\label{subsec:phaseTransition}Phase transitions in materials}

Phase transitions at surfaces and in small volumes play a key role in many systems, ranging from energy applications to biology. However, probing molecular dynamics with nanoscale spatial resolution is not possible using bulk measurement techniques. Quantum sensors can be useful in quantitative studies of such highly localized phenomena.

Understanding the phase transitions of liquid crystals is critical to utilizing them in display technologies and soft-matter applications. While studying these processes via conventional NMR is difficult due to the thin film nature of such devices, NV centers enable nanoscale NMR spectroscopy of liquid crystal phases. For example, Ref.~\cite{kavatamane_probing_2019} performed temperature-dependent proton NMR with single NVs to observe an increase in the diffusion constant of (5nm)$^3$ volume of the liquid crystal 8CB with temperature, indicative of different phase ordering. An even clearer signature of the liquid crystal phase transition is the motional narrowing of the NMR peak across the order-disorder phase transition. Ref.~\cite{pillai_observing_2026} observed this motional narrowing for the liquid crystal 5CB using shallow ensembles of NVs, as well as for solid-state barocaloric materials, capturing different thermal hysteresis behavior for the surface proximal bilayers (first $\sim$12~nm) vs the bulk of the material. 

Quantum sensors have also been utilized to study phase transitions in inorganic systems. The integration of NV sensing with diamond anvil cells (DACs) enables magnetometry with high sensitivity and resolution at pressures up to 140 GPa --- a regime where many conventional measurement techniques fail~\cite{hsieh_imaging_2019, steele_optically_2017, lesik_magnetic_2019, yip_measuring_2019}. Refs.~\cite{hsieh_imaging_2019, lesik_magnetic_2019} utilized a sample-proximal layer of NV centers in a DAC to measure the $\alpha$ → $\epsilon$ structural phase transition of iron (\textbf{Figure~\ref{fig: 4}b}). Using ODMR, they extracted the magnetic dipole strength of iron and observed the pressure dependent hysteresis indicating this bcc-to-hcp phase transition. Ref.~\cite{hsieh_imaging_2019} additionally used DC magnetometry with NVs in a DAC to map out the pressure-temperature phase diagram of gadolinium. They also demonstrated that T$_1$ relaxometry with nanodiamonds on a Gd film enabled detecting subtle structural and magnetic phase transitions, which would have been difficult to detect with DC magnetometry alone because of the minute changes in static magnetization.

NV magnetometry with DACs is also well-suited to studying high-T$_C$ superconductivity that arises from pressure-induced modification of microscopic interactions. By ODMR measurements, one can image local diamagnetism due to Meissner effect, thus directly probing the onset and properties of superconductivity across a wide range of parameters, including pressure, temperature, magnetic field, and spatial variations~\cite{lesik_magnetic_2019, yip_measuring_2019, bhattacharyya_imaging_2024}. Furthermore, by corroborating magnetometry with global electric transport, in situ local stress tensor mapping (also by NV centers), and stoichiometry mapping, Ref.~\cite{mandyam_uncovering_2026} obtained multi-parameter phase diagrams that show how these parameters locally affect superconductivity in the nickelate La$_3$Ni$_2$O$_7$ (\textbf{Figure~\ref{fig: 4}c}).

\subsection{\label{subsec:electrolyte}Electrolytes and batteries}

The non-destructive evaluation of intact battery cells is crucial for optimizing energy storage performance and ensuring operational safety. OPMs and NV centers have emerged as powerful complementary diagnostic tools which can detect external magnetic signatures to reveal internal battery dynamics. On the millimeter scale, OPMs enable magnetic imaging of current density distributions within active lithium-ion cells~\cite{evans_quantum_2025}. This magnetometric approach successfully uncovers hidden spatial inhomogeneities and weak transient internal currents indicative of early-stage degradation or defect formation~\cite{hu_sensitive_2020}. The rapid, completely non-invasive nature of OPMs facilitates scalable diagnostics for the continuous monitoring of emerging systems like solid-state batteries~\cite{hu_rapid_2020}. Complementing this macroscopic current mapping, NV sensing provides mesoscopic structural evaluation at the sub-millimeter scale. By exploiting the cross-relaxation feature between NVs and P1 centers at a 51.2 mT bias field, Ref.~\cite{zhang_battery_2021} demonstrated microwave-free eddy-current imaging of solid-state batteries. This technique achieved a spatial resolution of 360~\textmu m, successfully mapping the battery's architecture and identifying internal structural anomalies, such as brass impurities and crevices, without physical intrusion.

Beyond structural and current mapping, these quantum sensors provide unprecedented access to the chemical environment of battery electrolytes across vastly different length scales. For bulk analysis, OPM-detected ZULF NMR overcomes the fundamental limitations of traditional high-field NMR, where high-frequency radio waves are blocked by conductive metal casings. Low-frequency ZULF signals naturally bypass this skin effect. As a result, OPMs can directly monitor liquid electrolyte composition, degradation, and mass transport inside intact metal-cased batteries~\cite{fabricant_enabling_2025} (\textbf{Figure~\ref{fig: 4}d}). Conversely, at the microscopic solid-liquid interface, NV centers offer highly localized electrolyte sensing. Utilizing a microfluidic platform on a bulk oxidized diamond, Ref.~\cite{freire-moschovitis_role_2023} demonstrated that near-surface NV T$_1$ relaxation times actually increase in the presence of millimolar concentrations of diamagnetic electrolytes, such as aqueous NaCl. Diamagnetic ions alter interfacial band bending, which stabilizes charge fluctuations and reduces local magnetic and electric noise, thereby establishing a novel pathway for probing local electrolyte dynamics.

Finally, pushing quantum sensing directly to the active particle level, NV centers can be embedded within battery electrodes themselves to capture localized electrochemical phenomena invisible to macroscopic measurements. Ref.~\cite{liu_operando_2024} achieved~\textit{operando} quantum sensing by integrating nanodiamonds into the $\text{Fe}_3\text{O}_4$ electrode of a lithium cell (\textbf{Figure~\ref{fig: 4}e}). By tracking the highly localized magnetic stray fields produced during discharge, these embedded sensors mapped the nanoscale electrochemical phase evolution from $\text{Fe}_3\text{O}_4$ to $\text{FeO}$ and finally to $\text{Fe}$. This multi-threaded monitoring revealed significant spatial heterogeneity in reaction kinetics among individual active particles, capturing transient phenomena such as the superparamagnetic behavior of 5 nm reduced iron nanoparticles.

\section{\label{sec:sec6}Outlook}

\begin{figure*}[btp!]
\begin{center}
\includegraphics[width=\textwidth]{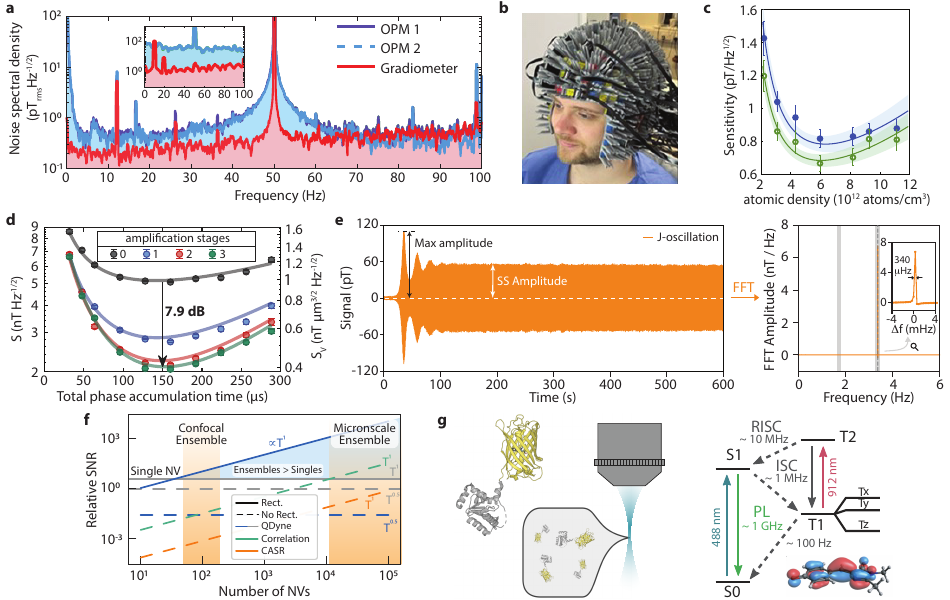}
\caption{{\bf{Advancements in quantum sensing for chemical applications}}
(a) Gradiometric configuration for enhanced OPM performance. Magnetic sensitivity of two OPMs operating in a differential gradiometric configuration is compared under unshielded environmental conditions. The gradiometric approach significantly improves noise rejection, enabling quantum sensing in demanding settings. Adapted from Ref.~\cite{xiao_movable_2023}. 
(b) Large-scale OPM system demonstrating the scalability of quantum sensing platforms. Adapted from Ref.~\cite{hill_optimising_2024}.
(c) Quantum-enhanced metrology with OPMs. Magnetic sensitivity is shown as a function of atomic density for coherent (blue) and squeezed (green) probe light, illustrating sensitivity enhancement achieved with quantum resources. Adapted from Ref.~\cite{troullinou_quantum-enhanced_2023}. 
(d) Interaction-enhanced magnetic sensing with an ensemble of NVs. Engineered many-body dynamics improve magnetic sensitivity through 1-, 2-, and 3-stage signal amplification. Adapted from Ref.~\cite{gao_nanoscale_2026}. 
(e) Sustained ZULF NMR signal oscillations in a J-oscillator driven by digital feedback and enabled by a continuous polarization source (left). Resulting NMR spectra exhibit sub-mHz linewidths (right). Adapted from Ref.~\cite{xu_quantum_2026}.
(f) Towards chemical shift resolution noise spectroscopy for nanoscale NMR. Phase rectification utilizing the nitrogen nuclear spin of the NV center allows for applying coherently averaged readout to the measurement of statistically polarized nuclear spins. Schematic shows how significantly higher SNR and spectral resolution ensemble NV-NMR can be achieved using such protocols. Adapted from Ref.~\cite{maier_efficient_2025}.
(g) Novel quantum sensors based on fluorescent proteins. Here, the EYFP sensor qubit is conjugated to a target glutaredoxin protein to be sensed, allowing for close sensor-sensee proximitization. Schematic of energy levels which allow for coherent manipulation and optical readout of the triplet metastable spin states (whose orbitals are shown). Adapted from Ref.~\cite{feder_fluorescent-protein_2025}. 
}
\label{fig: 5}
\end{center}
\end{figure*}

As discussed in Sections~3-5, quantum sensors offer several advantages over conventional spectroscopic techniques, allowing chemical and materials analysis in unprecedented regimes. Here, we highlight recent work towards further improving the usability, sensitivity, and spectral resolution of OPMs, NV centers, and other novel sensing platforms. 

Broad OPM deployment requires overcoming reliance on magnetic shielding via pulsed-operation schemes to avoid ambient dephasing~\cite{limes_portable_2020} and gradiometric detection for common-mode noise cancellation~\cite{xiao_movable_2023} (\textbf{Figure~\ref{fig: 5}a}). Concurrently, scaling these standalone sensors into high-density multichannel arrays, inspired by magnetoencephalography~\cite{hill_optimising_2024} (\textbf{Figure~\ref{fig: 5}b}), promises to unlock highly parallelized, high-throughput chemical screening. For NV centers, poor photon collection efficiency is a technical bottleneck. This can be improved by engineering the local photonic environment with prefabricated nanostructures like nanopillars~\cite{kim_scalable_2025} and direct-bonded diamond membranes~\cite{guo_direct-bonded_2024} for seamless microfluidic integration. Overcoming these barriers has enabled magnetic DC sensitivities below 10 pT$/\sqrt{\mathrm{Hz}}$~\cite{barry_sensitive_2024, sekiguchi_diamond_2024, omar_human_2026}, sufficient to map subtle ionic currents and slow catalytic processes.

The absolute sensitivity of a quantum sensor is fundamentally constrained by spin projection noise. To push beyond this classical limit, the next generation of OPMs will increasingly rely on true quantum enhancements, most notably spin squeezing. By leveraging light-atom interactions to induce controlled entanglement among alkali atoms in the vapor cell, researchers can redistribute the quantum uncertainty of the sensing variable, effectively suppressing it below the standard quantum limit\footnote{Standard quantum limit (SQL): The theoretical noise floor set by the random, independent statistical fluctuations of non-interacting sensor spins.}. Foundational demonstrations of this spin-squeezed sensitivity~\cite{sewell_magnetic_2012} (\textbf{Figure~\ref{fig: 5}c}), combined with recent breakthroughs in sustaining these non-classical states at the high atomic densities necessary for practical magnetometry~\cite{troullinou_quantum-enhanced_2023}, establish a clear trajectory for the field. Ultimately, integrating these quantum-enhanced states into standard analytical OPM platforms will drive profound improvements in SNR, enabling the detection of more elusive chemical signatures. Solid-state quantum sensors are pursuing a parallel trajectory toward overcoming the standard quantum limit. For NV centers, recent implementations utilizing the long nuclear memory at high field have successfully pushed the optical readout of solid-state spin ensembles to the fundamental projection noise limit~\cite{maier_readout_2026}. Venturing beyond this limit requires quantum enhancement. In recent works~\cite{wu_spin_2025, gao_signal_2025}, dipole interactions between nearby spins were used to achieve such an enhancement through spin squeezing and signal amplification. While the metrological gain was severely limited in these initial works, a new mechanism~\cite{put_collective_2025, gao_dressed-state_2026} to generate metrologically useful dynamics was exploited in Ref.~\cite{gao_nanoscale_2026}. This yielded a substantial 7.9~dB improvement in AC magnetic field sensitivity, even after accounting for experimental overhead (\textbf{Figure~\ref{fig: 5}d}). While this already provides a practical sensitivity enhancement for state-of-the-art micro- and nanoscale magnetometry, combining this technique with high-fidelity readout, appropriate diamond surface terminations to mitigate magnetic noise, and robust surface functionalization to deterministically interface with molecular targets~\cite{janitz_diamond_2022} will further unlock the potential of NV sensing.

Maximizing spectral resolution is another critical frontier. For OPM-detected ZULF NMR, quantum magnetic J-oscillators~\cite{xu_quantum_2026} employ resonant feedback to sustain coherent spin dynamics, dramatically extending coherence times and narrowing spectral lines (\textbf{Figure~\ref{fig: 5}e}). Towards chemical shift resolved nanoscale NV-NMR, phase rectification has allowed for applying time-efficient coherent averaging techniques to incoherent nuclear spin noise~\cite{maier_efficient_2025} (\textbf{Figure~\ref{fig: 5}f}). Alternatively, translating macroscopic NV-ZULF techniques~\cite{omar_zero-_2026} to the nanoscale offers a transformative pathway for probing chemistry directly at the solid-liquid interface. Beyond single-point magnetometry, new fundamental modalities are emerging. Nanoscale covariance magnetometry\footnote{Covariance magnetometry: A measurement protocol that cross-correlates the readouts of multiple quantum sensors to isolate correlated magnetic noise from uncorrelated local environments, directly probing the two-point correlation function of a target system.}~now allows the extraction of spatiotemporal correlations suitable for complex chemical networks~\cite{rovny_nanoscale_2022, le_wideband_2025}. Concurrently, the NV center's sensitivity to strain and temperature enables the mapping of exothermic catalytic ``hot spots''~\cite{kucsko_nanometre-scale_2013}, and operation at gigapascal pressures in DACs could enable simulating geochemical processes~\cite{bhattacharyya_imaging_2024, casey_promise_2020, wang_optically_2021}. Both platforms are also positioned to probe molecular chirality, with proposals to use OPM zero-field NMR to discriminate enantiomers~\cite{king_antisymmetric_2017} and NVs to track chiral-induced spin selectivity~\cite{volker_toward_2023}.

Finally, the quantum sensing toolkit is expanding beyond traditional hosts to include new defects and molecular systems~\cite{vaidya_quantum_2023, luo_fabrication_2023, wasielewski_exploiting_2020, yu_molecular_2021}. Promising examples for chemistry applications include robust near-infrared silicon carbide defects for single-molecule NMR~\cite{chen_single-molecule_2025} and bottom-up design of surface-scaffolded molecular qubits on 2D materials to minimize sensor-sample distances~\cite{zheng_surface-scaffolded_2026, zhou_optically_2026}. Ultimately, leveraging biomolecular machinery, such as optically addressable spin qubits encoded in fluorescent proteins~\cite{feder_fluorescent-protein_2025} (\textbf{Figure~\ref{fig: 5}g}), promises to revolutionize targeted~\textit{in vivo} chemical sensing applications.

\section*{Disclosure statement}
M.D.L. and H.P. are shareholders of Quantum Diamond Technologies Inc..
The authors are not aware of any other affiliations, memberships, funding, or financial holdings that might be perceived as affecting the objectivity of this review.

\begin{acknowledgments}
The authors would like to thank Szymon Pustelny for his valuable feedback. This work was supported by the National Science Foundation (grant number PHY-2012023), the Center for Ultracold Atoms (an NSF Physics Frontiers Center), Gordon and Betty Moore Foundation (Grant No.~7797-01) and NSF QuBBE QLCI (NSF OMA-2121044). A.P. acknowledges support from DoD NDSEG Graduate Research Fellowship. 
\end{acknowledgments}

\bibliography{Bib_aps.bib}
\end{document}